\documentclass{article}

\usepackage{cite}
\usepackage[latin1]{inputenc}
\usepackage{slashed}
\usepackage{amsmath}
\usepackage{textcomp}
\usepackage{amssymb}
\usepackage{amsfonts}
\usepackage{indentfirst}
\usepackage{color}
\usepackage{graphicx}
\usepackage{placeins}
\usepackage{tikz}
\usepackage{hyperref}
\usetikzlibrary{backgrounds}
\tikzset{phi/.style ={color= blue!50}}
\tikzset{psi/.style ={color= red!50}}
\newcommand{\be}{\begin{equation}}
\newcommand{\ee}{\end{equation}}
\newcommand{\bea}{\begin{align}}
\newcommand{\eea}{\end{align}}

\newcommand{\nn}{\nonumber}

\newcommand{\ra}{\rangle}
\newcommand{\la}{\langle}
\newcommand{\Tr}{\mathrm{Tr}}

\begin{document}
\title{Perturbative calculation of field space entanglement entropy}
\author{James Brister\\
\normalsize\textit{Center for Theoretical Physics, College of Physical Science and Technology,}\\
        \normalsize\textit{Sichuan University, 29 Wangjiang Road, Chengdu 610064, P.~R.~China}\\
        \normalsize\textit{E-mail:}
        \texttt{jbrister@scu.edu.cn,}
}

\maketitle
\begin{abstract}
We present a general method for the perturbative calculation of the entanglement entropy between two interacting quantum fields. Previous attempts at calculating this quantity perturbatively have encountered a seemingly pathological divergence; we explain why this divergence is a result of improperly truncating a series expansion and give a prescription for avoiding this problem. We then apply our method to a simple example of two mass-mixing scalar fields.
\end{abstract}
\section{Introduction}

\subsection{Background}
In order to better understand the entanglement structure of quantum field theory, we would like to understand the nature of the entanglement entropy between any  partitions of the Hilbert space. In perturbative quantum field theory at least, the partition of Hilbert space into the separate configuration spaces of the various fields of the theory seems a natural choice; however, this quantity, called ``field'' or  ``field space'' entanglement entropy in some previous works, has been relatively under-studied. Most previous investigations have focused on the rare special cases in which exact solutions can be found or some other non-perturbative method used, perhaps because of a perceived difficulty in obtaining perturbative results, which appear to be necessarily divergent (even with appropriate cutoffs imposed). In this article we explain the origin of this divergence and show that it is readily avoided by a more careful approach to the interaction of series expansions with limits.\\

Holographic duals of the field entanglement have been proposed (generally being spacetimes with two interacting asymptotic regions, see \cite{Taylor:2015kda}) however we will not devote much attention to such considerations here as we believe the field entanglement entropy is a quantity of interest in itself for understanding the structure of quantum information in field theories, in broader circumstances than the CFTs for which a gravity dual can be defined. As a more practical application, consider the case of some readily observable fields coupled weakly to a hidden sector (such as in many dark matter models). One might consider the extent to which it is appropriate to treat the vacuum, say, of the full theory a pure state for the non-hidden fields alone, and what uncertainties this may introduce to our attempts to model these fields. \\

\subsection{Previous work}
An early example of attempts to calculate these quantities perturbatively is \cite{Teresi:2010kt}, wherein the authors used the replica trick to calculate the $n$th R\'enyi entropy for a pair of interacting scalar fields, could not obtain a finite result in the limit $n \rightarrow 1$. As explained in the next section, the problem here comes from truncating a series expansion \emph{before} taking the limit, when the order of some of the terms depends on $n$.\\
Another early work on the subject, \cite{Yamazaki:2013xva}, considered the entanglement between two sectors interacting through a gauge field, but did not make any explicit calculations. \\
The term ``field entanglement entropy'' is taken from  \cite{Taylor:2015kda}, in which it is considered as a special case of entanglement entropy between factors of Hilbert space related by symmetry groups.\\
In \cite{Mollabashi:2014qfa}, simple mixing models were considered, for which the ground states can be exactly calculated. Finite entanglement entropy densities were found for these models, but an attempt to recover these results from perturbation theory encountered the same problem as \cite{Teresi:2010kt}, though the authors correctly suggested that their divergent terms might be canceled by a term that did not appear at the same order in the series expansion. This work was later generalized to more fields\cite{Mozaffar:2015bda}.\\
A similar entropy was calculated in \cite{Boyanovsky:2018fxl}, by considering effective potentials for thermal QFTs in which more massive particles had been integrated out.\\
In \cite{Xu:2011gn} the R\'enyi entanglement entropies of interacting conformal field theories were considered,as well as those of interacting fermi liquids, and it was argued that perturbation theory at finite temperature produces the expected scaling behavior.\\
In \cite{Nakai:2017qos}, the thermodynamical effect of entanglement between decoupling fields in an expanding universe is considered, and an explicit prescription for calculating entanglement entropy at finite temperature is given. However, this paper is mostly concerned with the high-temperature limit, and it is unclear whether taking the $T \to 0$ limit makes sense, as the topology of the manifolds over which one must take path integrals changes in this limit.\\
 \cite{Polyakov:2019clc} investigated the entanglement between sectors of different spin in String Field Theory.

\subsection{Structure of this article}

In the next section we explain the origin of the divergence found by previous studies, and show how it results from an improper treatment of limits. We then derive the correct procedure to obtain perturbative contributions to the field entanglement entropy at lowest order in the coupling. 

In section 3 we apply this procedure to a simple example (a pair of scalar fields interacting through a mass-mixing) and obtain results compatible with the conclusions of other studies.\\

\section{General argument}

\subsection{Set up}
Consider two fields $\phi$,$\psi$, interacting through some polynomial term (perhaps with derivatives) $\mathcal{L}_{int}= g F(\phi, \psi)$, where $g$ is an interaction strength that we shall assume to be small. Given the vacuum state $|0\ra$, we write the density matrix as a Euclidean path integral

\be
\rho = |0\ra\la0| = \frac{1}{Z}\int D\phi_i D \psi_i D\phi_f D\psi_f |\phi_i\ra|\psi_i\ra\la\phi_f|\la\psi_f| \int D \phi D \psi e^{-S(\phi,\psi)}~,
\ee
where $Z$ is the partition function, the first set of integrals are over spatial configurations at \emph{equal time} ($\tau=0$, say) and the latter are over spacetime configurations with the boundary conditions $\phi(0^-,x)=\phi_i(x), \phi(0^+,x)=\phi_f(x)$ (and similarly for $\psi$).\\

We trace out one of the fields, $\psi$ say, to obtain a \emph{reduced density matrix}
\be
\rho_r = \int D \psi' \la\psi'| \rho |\psi'\ra =  \frac{1}{Z}\int D\phi_i D \psi' D\phi_f  |\phi_i\ra|\la\phi_f| \int D \phi D \psi e^{-S(\phi,\psi)}~,
\ee
where the second integral now has the boundary condition $\psi(0,x) = \psi'(x)$ (combined with the first integral this is, of course, equivalent to taking an integral over all possible configurations for $\psi(\tau,x)$, but we retain this form for the sake of clarity). \\
We define the entanglement entropy to be the entropy of this mixed state on $H_\phi$:
\be
S_{EE}= \Tr \rho_r \log(\rho_r)
\ee

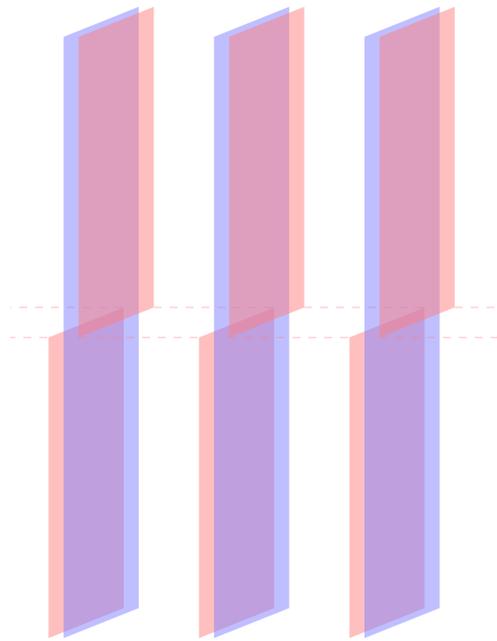
\begin{figure}

\begin{tikzpicture}
\clip (2.5,0) rectangle (9,8.5);
\def\psipath{-- +(1,0.4) -- +(1,4.4) -- +(0,4)}
\def\phipath{-- +(1,0.4) -- +(1,8.4) -- +(0,8)}
\foreach \x in {0,1,2,3,4}{
\begin{scope}[xshift= \x*2.0 cm]
	\fill[psi, semitransparent]  (1,0) \psipath ;
	\fill[phi, semitransparent] (1.2, 0) \phipath;
	\fill[psi, semitransparent]  (1.4, 4) \psipath  ;
	\draw[psi, semitransparent, dashed] (1.4,4) -- (3,4);
	\draw[psi, semitransparent, dashed] (2.4,4.4) -- (4,4.4);
\end{scope}
}
\end{tikzpicture}

\caption{Schematic view of the manifold $M_n$. A copy of one of the fields fields, $\phi$ say, lives on each of the blue sheets, while the copies of $\psi$ fields live on the red. For $\tau<0$, the $i$th $\phi$ couples to the $i$th $\psi$, but for $\tau>0$ it couples to the $i+1$st.}
\label{sheets}
\end{figure}

Calculating $S_{EE}$ from the above expression is, in general, rather difficult, so we employ the ``replica trick'':  we calculate the $n$th R\'enyi entropies
\be
S_n = \frac{1}{1-n}\log \Tr[\rho_r^n]
\ee
for arbitrary integers $n>1$, analytically continue, and take the limit $n \rightarrow 1$. Taking the product of two copies of $\rho_r$ amounts to imposing that $\phi_f$ for the first is equal to $\phi_i$ for the second, and taking the path integral over all such configurations for this $\phi(0,x)$. The quantity $\Tr[\rho^n]$ is thus
\be
\frac{1}{Z^n} \int_{M_n} D\phi D\psi e^{-S},
\ee
where $M_n$ is an unusual spacetime consisting of $n$ copies of the usual Euclidean spacetime: on each copy, the field $\psi$ propagates normally, however the field $\phi$ sees a cut at $\tau=0$ that joins the lower half of the $i$th copy to the upper half of the $i+1$st copy and, similarly, the $n$th copy to the first due to the trace (see figure \ref{sheets}). As should be expected, this situation is symmetric under interchanging $\phi$ and $\psi$.\\

 We shall write $Z_n$ for the partition function on $M_n$. We therefore need to calculate $\log \frac{Z_n}{Z^n}$, the logarithm of the ratio of the partition function on $M_n$ to that on $n$ copies of the usual euclidean spacetime (which we may call $M_1$). Following the argument of \cite{Teresi:2010kt} we see that this quantity can be written in terms of the generating functions for \emph{connected} Feynmann diagrams as 
\be
S_n = \frac{1}{n-1} ( n W- W_n)
\ee
Where $W_n$, $W$ are the generating functions for connected diagrams on $M_n$ and $M_1$, respectively.\\

\subsection{Method}
In terms of Feynman diagrams, the contributions to the R\'enyi entropies will come from connected vacuum diagrams that are allowed on either $M_n$ or the $n$ copies of $M_1$, but not both. These will necessarily always have propagators crossing from $\tau>0$ to $\tau<0$, we may impose these conditions by introducing Heaviside step functions $ \Theta(\tau),\Theta(-\tau')$ at each vertex.

 The simplest of these diagrams, correctly identified in previous work, is a two-vertex diagram in which propagators for \emph{both} fields connect one vertex at $\tau >0 $ and one at $\tau<0$, which may only happen on the $n$ copies of normal Euclidean space \footnote{Indeed, the only diagrams that may contribute are ones in which propagators of both fields connect vertices above and below $\tau=0$}. Schematically, if $D_\phi(x,t,,x',t')$ represents the $\phi$ propagators, and $D_\psi$ the $\psi$ (however many of each line there might be), this gives a contribution of 
\be
n \int_{\tau>0}d\tau d^dx\int_{\tau'<0}d\tau' d^dx' g^2 D_\phi(x, \tau, x', \tau') D_\psi(x, \tau, x', \tau') \equiv n A_1
\ee
to $Tr[\rho_r^n]$.\\

This contribution to the R\'enyi entropy clearly diverges as $n \rightarrow 1$ (as will any higher order terms derived from such diagrams) but there appear to be no other contributions at the same order in the coupling $g$. This is correct, but does not imply a divergence in the entanglement entropy: to have a consistent expansion in $g$ to some order for $S_{EE}$, we must include all those terms contributing to $S_n$ \emph{that become of that order in the $n \rightarrow 1$ limit}.\\

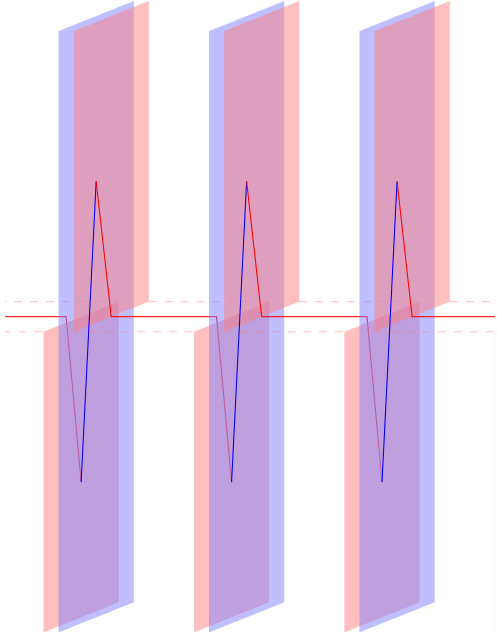
\begin{figure}[h]
\begin{tikzpicture}
\clip (2.5,0) rectangle (9,8.5);
\def\psipath{-- +(1,0.4) -- +(1,4.4) -- +(0,4)}
\def\phipath{-- +(1,0.4) -- +(1,8.4) -- +(0,8)}
\foreach \x in {0,1,2,3,4}{
\begin{scope}[xshift= \x*2.0 cm]
	\fill[psi, semitransparent]  (1,0) \psipath ;
	\fill[phi, semitransparent] (1.2, 0) \phipath;
	\fill[psi, semitransparent]  (1.4, 4) \psipath  ;
	\draw[psi, semitransparent, dashed] (1.4,4) -- (3,4);
	\draw[psi, semitransparent, dashed] (2.4,4.4) -- (4,4.4);
	\draw[blue] (1.5, 2) -- (1.7,6);
	\draw[red] (1.7,6) -- (1.9,4.2) -- (3.3,4.2) -- (3.5,2);
\end{scope}
}
\end{tikzpicture}

\caption{The process contributing to the $n$th R\'enyi entropy at order $g^{2n}$. Red (blue) lines represent propagators of fields on the red (blue) sheets.}
\label{loop}
\end{figure}

 At order $g^2$, there is therefore one more term we must consider: the diagram with $2n$ vertices alternating between $\tau >0$ and $<0$ with alternating $\phi$ and $\psi$ propagators joining each vertex to the next, clearly this is possible on $M_1$ but, as shown in figure \ref{loop}, it is also found on $M_n$. For consistency, we must subtract this diagram from power series of diagrams on $M_1$.\\
\FloatBarrier

\FloatBarrier

Schematically, this diagram contributes
\be
g^2n  \prod_{i=1}^n \int_{\tau_i>0}d\tau_i d^dx_i\int_{\tau_i'<0}d\tau' d^dx_i' g^2 D_\phi(x_i, \tau_i, x'_i, \tau'_i) D_\psi(x'_i, \tau'_i, x_{i+1}, \tau_{i+1}) = A_2(n)
\ee
with the opposite sign to the previous contribution. Taking a Taylor series about $n=1$, we see 
\be
A_2(n) = A_1 + (n-1) g^2 Tr[ D_\phi \Theta D_\psi \Theta \log(g^2 D_\phi \Theta D_\psi \Theta)] +O((n-1)^2)~,
\ee
where we have formally treated the propagators as linear operators.
\FloatBarrier
 The first term removes the previously observed divergence, by changing the factor of $n$ to a factor of $n-1$, canceling the denominator so the $A_1$ contribution is, in fact, independent of $n$. The second term gives a contribution to the entanglement entropy that is also of order $g^2$, plus a logarithmic term proportional to $A_1$.\\

We again emphasize that the only way to obtain a consistent perturbative expansion is to truncate the series \emph{only after taking any limits that may change the powers of the coupling}.

Having found a consistent way to perform this expansion, we explicitly calculate the field entanglement entropy for a relatively simple example in order to demonstrate the calculation techniques needed explicitly evaluate the contribution to the entropy derived above.

\section{Example: mass-mixing scalars}

We will consider the case of a ``two-point'' interaction of a pair of scalars coupled through an non-diagonal mass matrix. We choose this example for two reasons: firstly to minimize the number of momenta we must integrate over and secondly because this entanglement entropy has previously been approximated through non-perturbative means in \cite{Mollabashi:2014qfa}, giving us a check on the consistency of our answers. We emphasize that the methods used here are not peculiar to this model, and could easily be adapted to any other perturbative quantum field theory.\\

We take the Lagrangian density to be:
\be
\mathcal{L} = -\phi_i\partial^2\phi_i - \phi_i M_{ij}\phi_j
\ee
where $\phi_i = (\phi_1, \phi_2)$ are two scalars and the mass matrix
 \be
M = \begin{pmatrix} m_1^2 & \frac{1}{2}g\\ \frac{1}{2}g & m_2^2 \end{pmatrix}
\ee
We assume $m_2^2 >m_1^2 \gg g $.\\

We shall treat the off-diagonal parts of the mass matrix as an interaction term, i.e.
\be
\mathcal{L} = -\phi_1(\partial^2 + m_1^2)\phi_1 - \phi_2(\partial^2 + m_2^2)\phi_2 -g \phi_1\phi_2~,
\ee
where $g$ has units of mass squared. (To treat $g$ perturbatively, we should properly non-dimensionalize it by factorizing out some mass scale, but as there are several mass scales in the problem we shall refrain from doing so for now).\\

We therefore want to calculate the contribution from the diagram shown in figure \ref{loop} (and the equivalent contribution with the $\tau>,< 0$ reversed). We may reduce this to just an integral over Fourier space by replacing the conditions $\tau>0$ with Heaviside step functions $\Theta(\tau)$ which may be represented by 
\be
 \Theta(\tau)= \int dP \frac{e^{iP \tau}}{(2\pi) i(P - i \epsilon)},
\ee
in the limit as $\epsilon$ goes to zero from above. This Cauchy principal value for the Fourier transform of the Heaviside function is significantly easier to work with in this situation than the often-quoted $\frac{1}{ i P}+ 2\pi \delta(P)$.\\

We write $D_1, D_2$ for the propagators of the corresponding fields. To avoid an excess of indices, we will denote the (Euclidean) time components of spacetime vectors with capital letters $P,Q,R, \ldots$ and the space components by the corresponding lower case letters.
We therefore wish to calculate the contribution to $S_n$ of the diagram with $2n$ vertices at points $x_i, y_i~ i=1\ldots n$, such that $X_i <0$ and $Y_i>0$. 
\begin{align}
I_n &= g^{2n} \int \prod_{i=1}^n \int d^{d}x_i dX_i d^{d}y_i dY_i D_1(x_i,X_i, y_i,Y_i)\Theta(Y_i)D_2(y_i,Y_i, x_{i+1},X_{i+1})\Theta(-X_{i+1}) \nn \\
&+[X>0, Y<0] \nn\\
&= g^{2n}\prod_{i=1}^n \int d^{d}x_i dX_i d^{d}y_i dY_i \int d^dp_i dP_i d^dq_i DQ_i dR_i dS_i \frac{1}{(2\pi)^{2d+4}}\nn \\
&\frac{-e^{i((x_i-y_i)p_i + (X_i-Y_i)P_i +Y_iR_i +(y_i - x_{i+1})q_i (Y_i -X_{i+1})Q_i + X_{i}S_i)}}{(P_i^2 +p_i^2+m_1^2)(R_i-i\epsilon)(Q_i^2 +q_i^2+m_2^2)(-S_i-i\epsilon)} +[X>0, Y<0]
\end{align}
where the last term represents the similar contribution from the case $Y_i <0, X_i >0$.

Firstly, we can perform all but one of the spatial integrals as usual, setting all the $p_i, q_i$ to a single loop momentum $p$ and obtaining an overall $d$-dimensional spatial volume factor $V$. Similarly, we preform the $X_i$ and $Y_i$ integrals to obtain delta functions imposing $R_i= P_i-Q_i$ and $S_i = Q_{i-1}-P_i$. We now have
\begin{align}
I_n&= g^{2n}\frac{V}{(2\pi)^{n(2d+4)}} \int d^d p  \prod_{i=1}^n \int dP_i dQ_i\nn\\& \frac{1}{(P_i^2+p^2 +m_1^2)(Q_i^2+p^2+m_2^2)(P_i-Q_i-i\epsilon)(Q_{i-1}-P_i+i\epsilon)} \nn \\&+[X>0, Y<0] ~.
\end{align}
We treat the $P_i$ integrals as contour integrals, and close them in the lower half-plane, picking up the contribution from the pole at $-i\sqrt{p^2+m_1^2}$. We now have
\begin{align}
I_n =& g^{2n} \frac{V}{(2\pi)^{n(2d+3)}} \frac{1}{(-2)^n} \int d^d p \frac{1}{(p^2+m_1^2)^\frac{n}{2}} \prod_{i=1}^n\int dQ_i \frac{1}{(Q_i^2+p^2+m_2^2)(Q_i + i\sqrt{p^2+m_1^2})^2} \nn\\&+[X>0, Y<0]~.
\end{align}
Similarly, we close the $Q_i$ contours in the upper half-plane to obtain
\be
I_n =2V \int d^d p\left[ \frac{g^2}{4(2\pi)^{d+1}} \frac{1}{\sqrt{p^2+m_1^2} \sqrt{p^2+m_2^2}(\sqrt{p^2+m_1^2}+\sqrt{p^2+m_1^2})^2}\right]^n
\ee

where we introduce a factor of 2 to account for the $Y_i <0, X_i >0$ case\footnote{we close the integration contours in the opposite directions, picking up an even number of overall factors of $-1$.}.\\
 At this point, it is easy to see that this quantity is indeed dimensionless and to factor out some arbitrary mass scale from $p, m_i, g, V$. We shall continue to use the same symbols for these quantities, but from here on consider them to be dimensionless numbers and note that the answer should only depend on ratios of the massive quantities $g, m_i^2$ (the volume factor compensates for the scaling of the $d$-dimensional momentum integral).\\

Performing this integral for arbitrary $d$ presents some difficulties, but the case $d=2$ is more tractable. We first take a series about $n=1$ to obtain
\begin{align}
I_n &=& 2 V \int d^d p \frac{g^2}{32\pi^3}\bigg[  \frac{1}{\sqrt{p^2+m_1^2} \sqrt{p^2+m_2^2}(\sqrt{p^2+m_1^2}+\sqrt{p^2+m_1^2})^2}  \nonumber\\ 
&+&\frac{(n-1)\log \left(\frac{g^2}{32\pi^3}   \frac{1}{\sqrt{p^2+m_1^2} \sqrt{p^2+m_2^2}(\sqrt{p^2+m_1^2}+\sqrt{p^2+m_1^2})^2} \right) }{\sqrt{p^2+m_1^2} \sqrt{p^2+m_2^2}(\sqrt{p^2+m_1^2}+\sqrt{p^2+m_1^2})^2} \bigg] \nonumber \\
&+&O(n-1)^2
\end{align}

The first term, as expected, is equal to $\frac{-1}{n}$ times the the contribution from the $n$ single sheets.
 We can perform the integral to find the lowest order contribution to the entropy

\begin{align}
S_1 =& V \frac{g^2}{16\pi^2}\bigg(\frac{1}{(m_1+m_2)^2}  \nn \\
+& \frac{m_1^2\log(\frac{32\pi^3 m_1^2 (m_1+m_2)^2}{g^2}) +m_2^2\log(\frac{32\pi^3 m_2^2 (m_1+m_2)^2}{g^2})- 2m_1 m_2 \log (\frac{32\pi^3 m_1 m_2 (m_1+m_2)^2}{g^2})}{(m_1^2-m_2^2)^2} \bigg)\nn \\
+& O(g^3)
\end{align}
As expected, only dimensionless combinations of the $m_i$, $g$  and $V$ appear in the answer. Treating $g$ as dimensionless, we can easily factor out the $\log g^2$ contribution to the second term above as $\frac{-\log(g^2)}{(m_1+m_2)^2}$, proportional to the first term.

\subsection{Discussion}
For some ``realistic'' values of the masses and couplings, $m_1 = 1, m_2 = 10, g=0.1 $, we find $S/V \sim 10^{-5} $, which is appropriately small. Indeed, the only way we may obtain $S/V \sim 1$ or larger for $g<1$ is by approaching the line $m_1=m_2$, where the assumptions used in this derivation fail to hold.\\

In \cite{Mollabashi:2014qfa}, this entanglement entropy was calculated using non-perturbative methods,  and for $d=2$ an approximation to the exact result found was, in our notation, proportional to $\frac{-g^2 \log g^2}{(m_1+m_2)^2}$, the same form as the leading logarithmic contribution we found above. \\

The most notable feature of this result is that it is \emph{finite} (per unit volume)\footnote{For other models, the field entanglement entropy should naively have the same divergence as the two-vertex vacuum graph that is its leading contribution}, and should remain so for $d<4$. That the entropy scales with volume is to be expected: unlike the more commonly studied spatial entanglement entropy, in which the leading contribution is from short-range interactions across the surface separating two regions, here the ``surface'' of interaction is the whole of $d$-dimensional space. \cite{Xu:2011gn} finds the same scaling in conformal field theories.

\section{Outlook and conclusions}

We have demonstrated that it is indeed possible to find the entanglement entropy between quantum fields by purely perturbative means. Moreover, our methods are readily generalized to other perturbative models, with $n$-point interactions leading to more complicated, but not inherently more difficult, integrals. Similarly, there would be little difficulty in extending this calculation to higher orders in the coupling or cases with background fields, so long as one is careful to include \emph{all} relevant diagrams. This opens up the possibility of calculating such quantities in any perturbative model - this may be of interest in studies of dark matter models, sterile neutrinos or any other model with a ``hidden sector''. \\

On a more technical note, we wish again to highlight the major point that previous studies had overlooked: in employing the replica trick in a perturbative setting, or any other situation where both a Taylor series and a limit are taken, one must be aware of the possibility of contributions whose order is a function of the variable for which the limit is being taken. As future studies continue to investigate entanglement between a wider variety of subsystems, it is likely that this point will prove relevant elsewhere.

\end{document}